\documentclass[%
 aip,
 amsmath,amssymb,
preprint,%
]{revtex4-1}

\usepackage{graphicx}
\usepackage{dcolumn}
\usepackage{bm}

\usepackage[utf8]{inputenc}
\usepackage[T1]{fontenc}
\usepackage{mathptmx}
\usepackage{etoolbox}
\usepackage{amsmath}
\usepackage[version=3]{mhchem}
\usepackage[capitalize]{cleveref}
\usepackage{xcolor}
\usepackage{booktabs}
\usepackage{lineno}
\usepackage{float}

\makeatletter
\def\@email#1#2{%
 \endgroup
 \patchcmd{\titleblock@produce}
  {\frontmatter@RRAPformat}
  {\frontmatter@RRAPformat{\produce@RRAP{*#1\href{mailto:#2}{#2}}}\frontmatter@RRAPformat}
  {}{}
}%
\makeatother

\begin{document}

\preprint{AIP/123-QED}

\title[Automatic graph representation algorithm]{Automatic graph representation algorithm  for heterogeneous catalysis}
\author{Zachary Gariepy}
\affiliation{ 
Department of Materials Science and Engineering, University of Toronto,  Toronto, Ontario M5S 3E4,Canada}

\author{ZhiWen Chen}
\affiliation{ 
Department of Materials Science and Engineering, University of Toronto,  Toronto, Ontario M5S 3E4,Canada}

\affiliation{ 
Department of Mechanical and Industrial and Engineering, University of Toronto,  Toronto, Ontario M5S 3E4,Canada}

\author{Isaac Tamblyn}%
\affiliation{ Department of Physics , University of Ottawa, Ottawa,Ontario  M5S 3G8, Canada
}%

\affiliation{Department of Physics , University of Ontario Institute of Technology, Oshawa, Ontario K1N 6N5,Canada}

\author{Chandra Veer Singh}
\affiliation{ 
Department of Materials Science and Engineering, University of Toronto,  Toronto, Ontario M5S 3E4,Canada}
\author{Conrard Giresse Tetsassi Feugmo}%
\affiliation{Department of Chemistry, University of Waterloo, 200 University Ave. West, Waterloo, Ontario N2L\,3G1, Canada; }

\email{cgtetsas@uwaterloo.ca}

\date{\today}

\begin{abstract}
\textcolor{black}{
One of the most appealing aspects of machine learning for material design is its high throughput exploration of chemical spaces, but to reach the ceiling of ML-aided exploration, more than current model architectures and processing algorithms are required. New architectures such as Graph Neural Networks (GNNs) have seen significant research investments recently.  For heterogeneous catalysis, defining substrate intramolecular bonds and adsorbate/substrate intermolecular bonds is a time-consuming and challenging process. Before applying a model, dataset pre-processing, node/bond descriptor design, and specific model constraints have to be considered. In this work, a framework designed to solve these issues is presented in the form of an automatic graph representation algorithm (AGRA) tool to extract the local chemical environment of metallic surface adsorption sites is presented. This tool is able to gather multiple adsorption geometry datasets composed of different systems and combine them into a single model. To show AGRA's excellent transferability and reduced computational cost compared to other graph representation methods, it was applied to 5 different catalytic reaction datasets and benchmarked against the Open Catalyst Projects (OCP) graph representation method. The two ORR datasets with O/OH adsorbates obtained 0.053 eV RMSD when combined together, whereas the three CO2RR datasets with CHO/CO/COOH obtained an average performance of 0.088 eV RMSD. To further display the algorithm's versatility and extrapolation ability, a model was trained on a subset combination of all 5 datasets with an RMSD of 0.105 eV. This universal model was then used to predict a wide range of adsorption energies and an entirely new ORR catalyst system and then verified through Density Functional Theory calculations.}
\end{abstract}

\maketitle

\section{Introduction}
Catalyzed energy reactions have emerged as a promising long term solution to a sustainable closed energy loop, however current catalysts do not meet the requirements for it to become a reality \cite{xin2020}. Current catalysts are not selective enough, and the few that show sufficient selectivity lack activity \cite{xin2020,pedersen2020}. This is the case for almost every electrocatalytic material group aside for high entropy alloys (HEAs), which have been shown to outperform current benchmark catalyst systems by up to 100\% selectivity and 200\% overpotential cost \cite{loffler2021,wu2020}. Due to the expansive configurational space of HEAs, their potential is incredible but largely untapped \cite{xin2020}. This material space is so large that even with the most efficient high throughput experimental methods, it would take years to find optimal geometries for a catalytic application. With the help of AI and computational modelling however, these configurational spaces can be thoroughly explored at a fraction of the time, monetary cost, and human effort \cite{xin2020,yan2021}. For example, Batchelor et al. explored binary catalyst configurations through a combination of Density Functional Theory (DFT) and machine learning (ML) for the oxygen evolution reaction (OER) and through optimization techniques, was able to propose the ideal catalytic composition of IrPt alloys \cite{batchelor2019}. Ma et. al took this a step further and utilized a similar ML approach to not only optimize a catalytic surface, but also design it in such a way that a specific reaction pathway in the Carbon Dioxide reduction reaction (CO2RR) was prioritized in order to produce long carbon chain products as a source of hydrocarbon fuel \cite{ma2015}. \\
Although Batchelor and Ma have already shown the merits of AI as a tool in catalyst design, there is still room for improvement. Current regression models mean absolute error (MAE) accuracy's when predicting trajectories, adsorption energies and other electronic properties lie on the range of 0.05-0.5 eV MAE \cite{batchelor2019,ma2015,chen2021,nayak2019,lu2020} and recent advancements in model complexities and spatial representation descriptors have been proven to be capable of pushing these accuracy's even further \cite{lu2020,chen_lu2021}. One particularly notable advancement was the transition from neural networks (NN) to graph neural networks (GNN), which can develop understanding of a local chemical environments configurational composition, not just chemical composition \cite{xie2018}. Additionally, traditional ML frameworks along with tree models and gaussian processes are not robust, require exact formatting of input vectors, and cannot be cross-trained on multiple adsorbates easily. The flexible input of graph neural networks can help bridge this gap between various incompatible models while simultaneously improving model performance. Back et Al., proved this by applying a GNN to predict binding energies on CO and H on diverse surfaces and reached MAEs of 0.15 eV \citep{Back2019}. Their graph representation of the crystals included information on the local environments of each atom computed using a statistical analysis of Voronoi polyhedra around each site, successfully encapsulating this information in a simple fashion using only the solid angle. Although this approach added flexibility to the input structure, there are still clear limitations to the method when multidentate adsorbates are brought into question.\\
Additionally, before applying a model to each of these studies, dataset pre-processing, descriptor design, and specific model constraints had to be considered. These processes demand large time commitments, can be prone to human error and are difficult to incorporate into future works. Although scientific insight can be gained from each individual work, concatenating their work to generate even more comprehensive models would be ideal and there is currently no clear path towards achieving this. GNNs contribute flexibility and spatial descriptors but the combination of datasets, incorporation of unique descriptor representations such as solid angles and efficient comparison of model frameworks have yet to be addressed.\\
In this work, another evolution in machine learning (ML) aided material design frameworks is proposed to solve major inefficiencies with current methods. We designed a unified graph representation with improved model performance while simultaneously reducing node counts  and computational costs which can be trained on numerous adsorbates with different coordination numbers. The framework is highly accurate, transferable and offers previously unavailable processing capabilities to combine multiple works together for boosted extrapolation ranges beyond single adsorbate and singular material system prediction. Using this tool, the data processing and model application steps of computational ML catalyst studies are vastly accelerated with new surface analysis functionality. This method also lays the foundations to predict catalytic properties on the worlds largest materials informatics databases such as Materials project and OpenCatalyst at a fraction of the computational cost and programming effort. 

\section{Construction of the graph for the adsorption sites}

\begin{figure}[H]
\centering
\includegraphics[width=0.95\linewidth]{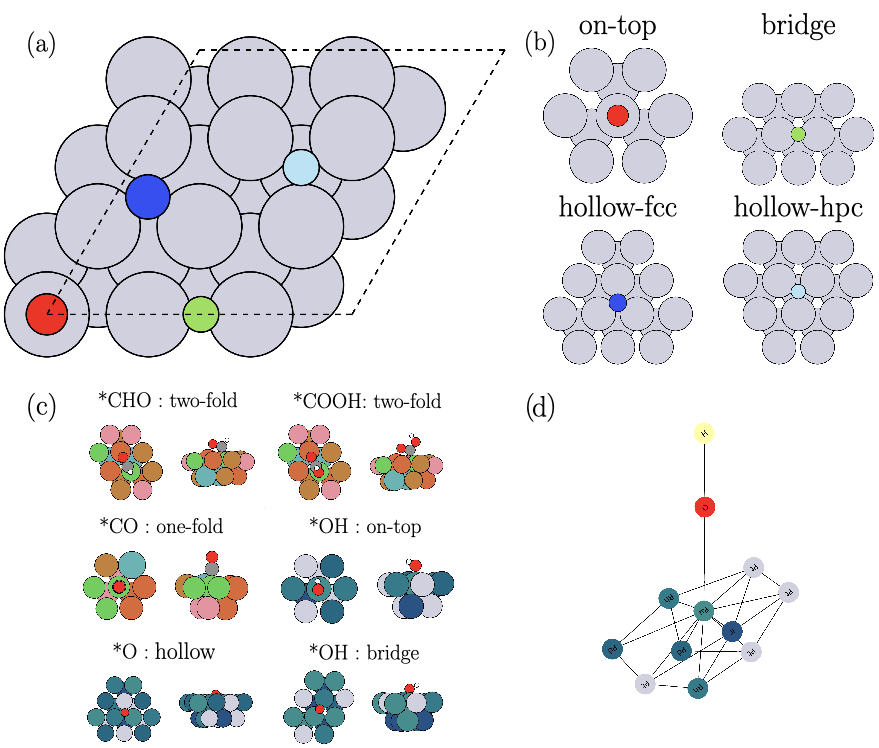}
\caption{ a) 4 AGRA recognized primary adsorption sites on a fcc (111) alloy surface. The color red, green, blue and light-blue correspond to on-top, bridge, hollow-fcc, and hollow-hpc sites. b) Generated local chemical environments extracted from AGRA surface analysis functionality. Periodic boundary conditions applied to each adsorption site in section (a) corresponding top sites (red), bridge sites (green), hollow-fcc (blue), and hollow-hpc (light-blue). c) Visualization of CO2RR and ORR HEA structures after adsorbate local chemical environment extraction is performed. d) conversion of the OH top-site local chemical environment geomtery from (c) into a GNN interpretable graph with atom/edge embedded descriptors}
\label{fig:fcc_111_ads_sites}
\end{figure}

Our graph representation of a material systems local chemical environment surrounding an adsorption site was built by a method inspired from Deshpande et. al. \cite{Deshpande2020}. The Python package Atomic Simulation Environment (ASE)\cite{Hjorth_Larsen_2017} was used to analyse a material systems surface and NetworkX \cite{hagberg2008exploring} was used to embed nodes and construct a graph representation of a given geometry file. A visual walkthrough of the algorithm can be seen in Figure S1. The algorithm's initial input is a geometry file of the adsorbate/catalyst system. Utilizing a user input specifying the desired adsorbate to analyze, the specified  molecule is identified and its indices are extracted from the input structures.\\
For each atom, the nearest neighbor atoms are defined using an atomic radius-based neighbor list generated with the ase.neighborlist module. This module used a radial cutoff for every atom based on metallic radii. A radius multiplier of 1.1 was also applied to all cutoffs for coarse grained adjustment and full encapsulation of awkwardly located adsorbate sites. Periodic boundary conditions are taken into account by unfolding bonds along the edge of the cell in repeats of the cell and two  atoms of the adsorbate are considered connected if the interatomic distances are lower than 1.8\;\AA. An atom of the adsorbate is considered connected to the catalyst (substrate) if the interatomic distance is lower than 2.3\;\AA. To be considered connected to a central atom, neighboring atoms must share a Voronoi face and have an interatomic distance lower than the sum of the Cordero covalent bond lengths. Once the nearest neighbors are extracted automatically, proximity based edge connection depending on if the node is an adsorbate or substrate atom is performed. For two adsorbate atoms, the cutoff is 2.3 \AA. For substrate atoms, the cutoff is 2.8 \AA. After identifying the catalyst atoms connected to the adsorbate, the algorithm selects their neighbors and removes the redundant atoms to generate a new structure of the local chemical environment surrounding the adsorbate. Redundant atoms are classified as duplicate atoms generated from periodic boundary conditions. As seen in \cref{fig:fcc_111_ads_sites} a-b, 4 primary types of adsorption sites can be identified. If 1 atom is considered connected to the adsorbate, the 'on-top' geometry will be extracted. if 2 atoms are connected, the bridge geometry will be extracted and if 3 atoms are connected, either hollow-fcc or hollow-hpc structures will be returned depending on the subsurface configuration of the adsorption site. Finally, the local chemical environment graph is generated using the extracted geometry where the nodes represent the atoms and the edges represent the relationship between neighboring atoms. This entire process only requires the user to input the adsorbate species of interest. Compared to other graph generation methods, neighbor radius cutoffs and edge cutoffs are fully automated. At each node, a feature vector is embedded following the procedure described in by Xie et al\cite{xie2018}. For the edges, the feature vector is constructed using the average pauling electronegativity between the atoms each node represented. Subsequently, the bond length represented with a Gaussian basis is appended. All gaussian basis parameters were taken from the CGCNN model and verified for consistent edge generation across the tested databases\cite{xie2018}. Through the use of basic JSON files, the node descriptors and edge attributes may be easily changed to test new configurations and spatial descriptors. This json interpretable atom embedding allows for the easy exploration and concatenation of new and previously discovered descriptors. The graph generation methodology is also applicable to a wide range of lattice orientations that extend beyond the fcc/hcp crystal structures highlighted in Section III.b. This allows for researchers to keep models basic for the study of parametric sensitivity towards descriptors or complex to capture complex intermingled phenoma influencing a catalytic site on a wide range of catalytic applications such as co-adsorption and asymmetric substrates. By default, a 16 or 92 descriptor set is applied to each model depending on the desired type of model \cite{xie2018,Chen2019}.\\
To further illustrate the surface analysis feature of AGRA, a visualization of each unique site geometry extracted from 2 publicly available databases is shown in \cref{fig:fcc_111_ads_sites} c. The GNN interpretable graph representation of an extracted site is also illustrated in \cref{fig:fcc_111_ads_sites} d. This geometry analysis feature is consistent because node count is not dependent on the initial size of the slab but only on the adsorption site type. \\
When compared to other graph representation methods such as the Open Catalyst Projects (OCP) atoms2graph function, AGRA's benefits are highlighted\cite{Chanussot2020}. The OCP generates its graph nodes by extracting a user specified amount of nearest neighbors (200 by default) without consideration for crystal structure orientation\cite{Chanussot2020}. Additionally, the node count of the OCPs graphing method can vary greatly if multiple databases with different simulation sizes are used. AGRA's framework extracts local chemical environments dependent on the surface crystal orientation around adsorbates to provide an additional layer of spatial description to the graph with greater node count consistency when considering multiple material systems. AGRA also further separates substrate atoms into binding site atoms and nearest neighbor atoms as opposed to OCP which separates substrate atoms into fixed (core) atoms and free (surface) atoms. Based on the databases discussed in Section 4, the AGRA graph representation yields superior results to the OCP graph representation when combining multiple adsorbate datasets which suggest the node generation method and local chemical environment extraction improve model transferability and versatility. The key differences are highlighted in Table I. For the technical limitations of AGRA regarding variable surface coverage and other situations, see Section A. in the Supplementary Information.

\begin{table}[!ht]
\caption{\label{tab:OCPvAGRA} Summarized differences between AGRA and OCP Graph Generation}
    \centering
    \begin{tabular}{lll}
    \hline
        Component\hspace{1.5cm} & \hspace{2cm}AGRA\hspace{2cm} & \hspace{3cm}OCP \\ \hline
        Node& Nodes based on adsorption geometry & Extracts N nearest neighbors from \\ 
        Generation & and local chemical environment & the input structure \\ 
        Edge & Edge connections based on & Fully connected nodes for all atoms \\ 
        Generation  & substrate adsorbate proximities & within specified distance of each other \\ \hline
    \end{tabular}
\end{table}

\section{Supervised learning to predict adsorption energy}
\subsection{The models}
Three graph neural networks were built on top of our representation to show the flexibility of our representation and its ability to add to existing frameworks. The first GNN will be referred to as 'NNConv' and consists of 3 major components : convolutional layers, recurrent layers, and pooling layers. These components are implemented in PyTorch Geometric \cite{Fey/Lenssen/2019}. The convolutional layers are continuous kernel-based convolutional operators \cite{DBLP:journals/corr/GilmerSRVD17}. To handle graphs of varying size and connectivity, a dynamic edge-conditioned filter \cite{DBLP:journals/corr/SimonovskyK17} was applied. The recurrent layer is a Gated Recurrent Unit (GRU) implemented in PyTorch. The pooling layer is the global pooling operator based on iterative content-based attention\cite{vinyals2016order}. First the nodes and edges are  embedded, then the convolutional and recurrent layers iteratively  update these features. After \emph{N} iterations, the pooling layer is then used for producing the overall feature vector.Finally, a fully-connected layer and output layer is used to predict the target property. The second GNN used is a crystal graph convolution neural network and will be referred to as 'CGCNN' \cite{xie2018}. This model consists of two major components: convolutional layers and pooling layers. The convolutional layers iteratively update the atom feature vector with surrounding atoms and bonds with a non-linear graph convolution function. After \emph{N} convolutions, the network automatically learns the feature vector for each atom. The pooling layer is then used for producing an overall feature vector. The third GNN used was an improvement on the second model published by Choudhary\cite{choud2022} and will be referred to as 'ALIGNN'. This model utilizes the  CGCNN framework but incorporates bond angle representation as well. The model updates nodes and edges via edge-gated graph convolution and sigmoid linear unit activations. After average pooling, the model runs through a final regression/classification layer to obtain a single value prediction. For the exact hyperparameters of each model used in training, see the Supplementary Information Section C.

\subsection{Results and discussion}

\subsubsection{ORR dataset} 

\begin{figure}[H]
\centering
\includegraphics[width=0.95\linewidth]{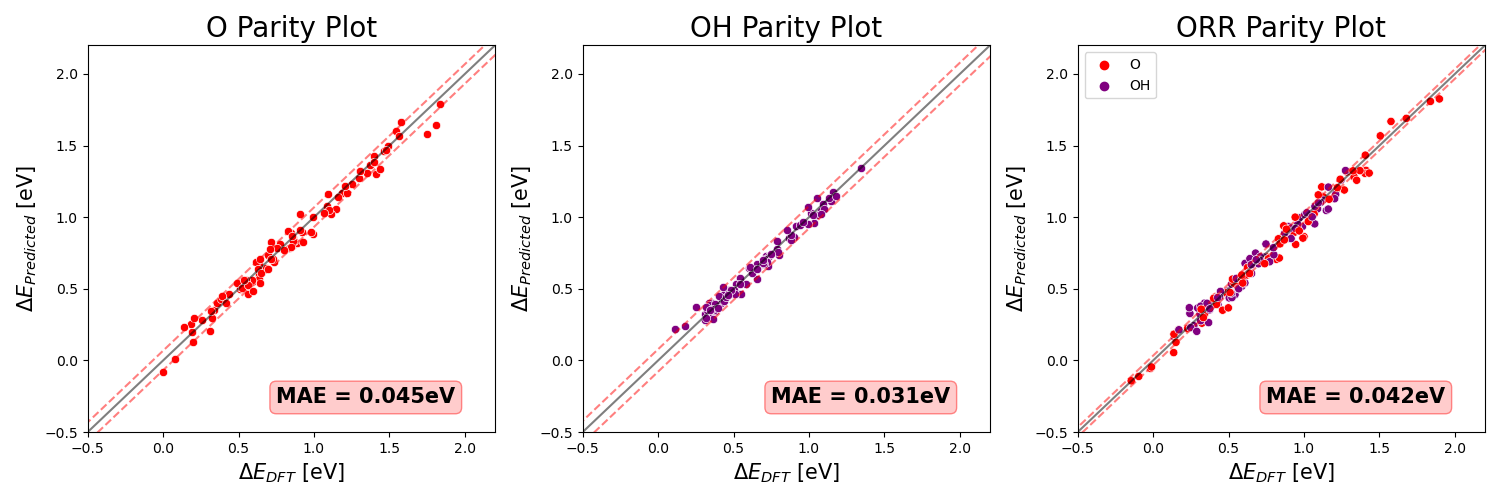}
\caption{Predicted adsorption energies plotted against DFT-calculated energies for the ORR test dataset using the top performing AGRA/ALIGNN model from 5 re-trains. MAE values highlighted were obtained from the most accurate saved model. O model was trained on 998 datapoints, OH model was trained on 871 and O/OH model was trained on 1600 datapoints.}
\label{fig:ORR}
\end{figure}

The first dataset we used to evaluate our graph representation was reported by T. Batchelor et. Al. \citep{batchelor2019}. This dataset deals with the oxygen reduction reaction (ORR) on \ce{RuIrRhPdPt} HEA. The calculations were performed using DFT through GPAW, an implementation of the projector-augmented wave method in ASE. The wave functions were expanded in plane-waves with an energy cuttoff of 400 eV and RPBE exchange-correlation functional. All slab calculation were performed with a minimum accuracy of 3x3x1 k-points Monkhorst-Pack sampling. HEA stability considerations used a 8x8x4 Monkhorst-Pack sampling.
The training set corresponds to the adsorption energies of *OH and *O at 871 and 998 different 2x2 unit cells, whereas the test set was modeled on 3x4 unit cells. This is a particular case where our approach could be applied because the size of the graph feeding the model does not depend on the size of the surface used for the simulation. \cref{fig:fcc_111_ads_sites}  c-d illustrates an example of extracted atoms for the *O adsorbed at the hollow sites and *OH adsorbed at the top and bridge sites, as well as their corresponding graph representation. Predicted adsorption energies plotted against DFT-calculated energies are also displayed in \cref{fig:ORR}. For each dataset, was more efficient and more accurate than that reported by  Batchelor  et. al. \cite{batchelor2019}. They obtained test set root-mean-square deviation (RMSD) of 0.063 and 0.076 eV for *OH and *O, respectively. Here, we obtained RMSD of 0.093 (*OH) and 0.172 (*O)  eV for  the NNConv model,  0.094 (*OH) and 0.149 (*O) eV  for the CGCNN model, 0.048 (*OH) and 0.059 (*O) for the ALIGNN model respectively. Each reported RMSE is the average of 5 train/val/test shuffles with 10/10/80 splits. Using the AGRA pipeline, each GNN was tested with a fractional amount of effort and time. When combining both datasets into a single model, an average RMSE of 0.053 eV was obtained, showing the transferability and flexiblity of the model. \\
For the combined dataset, 800 datapoints were randomnly selected from the O and OH datasets to create a single 1600 point combined ORR database. AGRA's ability to generate highly accurate models capable of predicting multiple reactions at high speeds with minimal processing has been effectively displayed with this dataset but its limitations are not tested with this dataset. This experiment only highlights the pipelines ability to predict on multiple adsorbates without sacrificing performance or speed. This is because the limited atom count of each datapoints graph converted local chemical environment dramatically reduces the computational cost of each GNN training epoch when compared to training with a radius cutoff based graph representation such as OCPs. 

\subsubsection{CO2RR dataset}

\begin{figure}[H]
\centering
\includegraphics[width=0.9\linewidth]{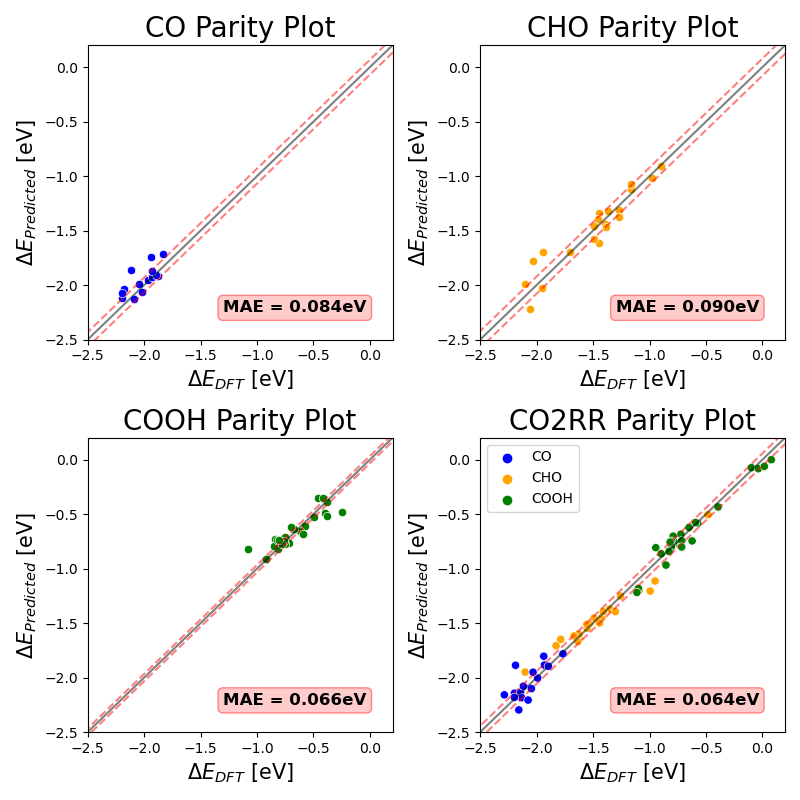}
\caption{Predicted adsorption energies plotted against DFT-calculated energies for the CO2RR test dataset using the top performing AGRA/ALIGNN model from 5 re-trains. MAE values highlighted were obtained from the most accurate saved model. CO,CHO, COOH models were trained on 170,204,267 datapoints. The combined CO2RR dataset was trained on 618 datapoints.}
\label{fig:CO2RR_parity}
\end{figure}

The CO2RR dataset taken from Chen et al. has 691 datapoints total spanning across CO,CHO and COOH adsorption energies on equiatomic CoCuFeMoNi high entropy alloys designed with a neural generator which maximizes system entropy \cite{Chen2022}. The same graph representation and nodal descriptors were used as in the ORR dataset. The HEA systems were (111) orientation, 64 atom slabs with adsorbates placed on every unique top site (16 top sites per slab). All DFT calculations for this database were conducted with the Vienna ab initio simulation package (VASP). Core electrons were described by the projector-augmented wave pseudopotential and the generalized gradient approximation (GGA) with Perdew-Burke-Ernzerhof functional was used. The wave function kinetic energy cutoff was 550 eV and 4x4x1 KPOINTs with a 15 \r{A}  vacuum gap. For symmetric adsorption molecules such as CO which are equally influenced from each of the 6 nearest surface neighbors and 3 subsurface neighbors, each top site was considered a single datapoint. For asymmetric adsorbates such as CHO and COOH, each top site yielded 6 unique datapoints. For atoms which are not directly bonded to the monodentate adsorption site, the van der Waals and electron cloud interactions of nearest neighbors were shown to influence these atoms. Because of this, each top site was calculated with 6 different orientations where the auxiliary atom was above the 6 surface nearest neighbors. \\
The authors reported MAE (eV) scores of 0.095, 0.095 and 0.068 eV for the CO,CHO and COOH datasets. Using AGRA in combination with the the ALIGNN and CGCNN models, we were able to achieve equivalent accuracy for the CO dataset, and superior accuracy for the CHO and COOH datasets. The ALIGNN model provided the lowest MAE scores for all three datasets. The CO dataset had an average MAE of 0.095 eV, CHO MAE of 0.095 eV and COOH MAE of 0.065 eV (\cref{fig:CO2RR_parity}). The strong accuracy of AGRA with minimal dataset preparation can be attributed to the scripts ability to accurately recognize bridge, hollow and top sites (\cref{fig:fcc_111_ads_sites} c) and automatically translate the local chemical environment into a descriptive readable graph for the GNN given the proper embedding descriptors for nodes. \\
Combining the 3 adsorbate datasets into 1 model resulted in a highly accurate model with an average MAE of 0.067 eV. Application of AGRA to this work highlights AGRAs versality when presented with multiple limited size datasets. Through a combination of adsorbate bond angle consideration and binding site edge representation, a singular model was capable of capturing not only the complex catalytic surface of HEAs which have been proven to break the linear scaling relation, but also consider the complexities of various adsorbate configurations. Chen's work utilized a multi-perceptron neural network to analyze the influence of the surrounding environment on each adsorbate dataset to ultimately show how HEAs break the linear scaling law to create superior catalysts. With AGRA, these same complexities were captured more accurately while simultaneously combining all 3 adsorbates into a singular database for even greater dynamics comprehension. 

\subsubsection{Combined CO2RR and ORR Dataset}

\begin{figure}[H]
\centering
\includegraphics[width=0.9\linewidth]{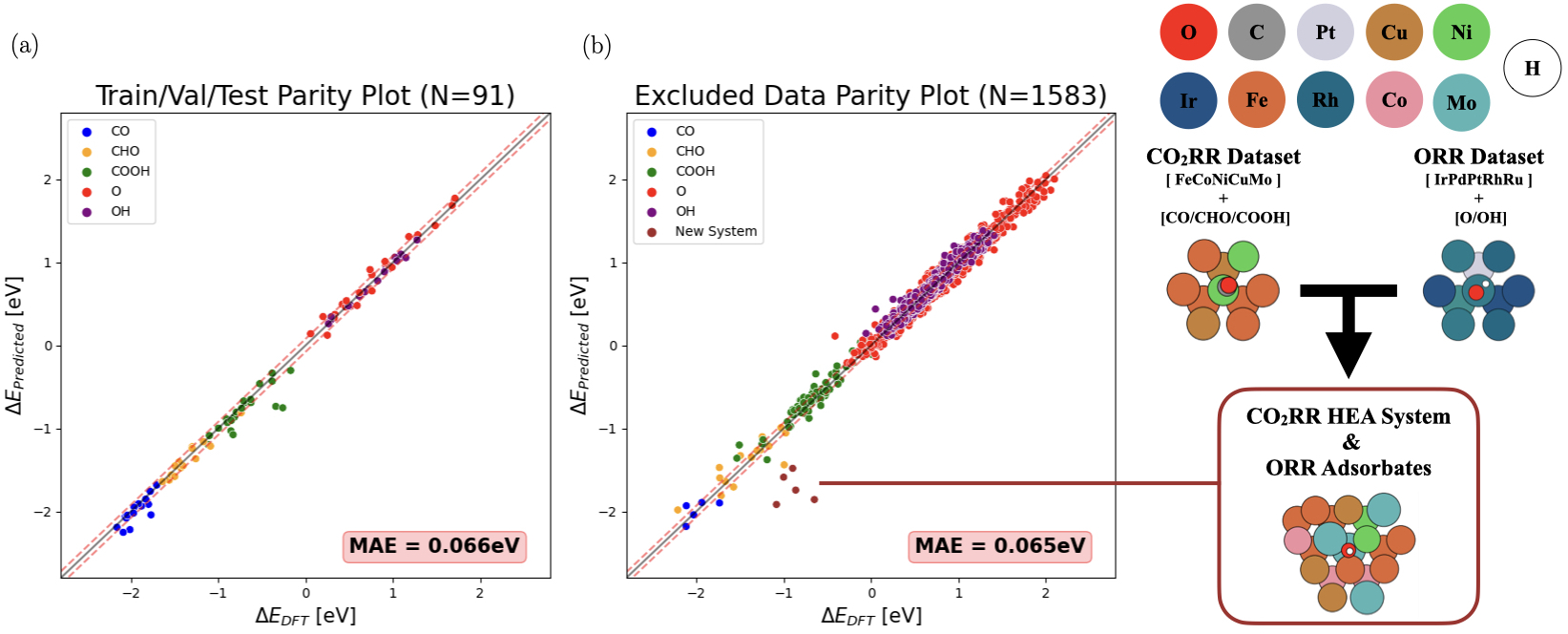}
\caption{a) Combined Dataset Parity Plot with 950 datapoints split between train/validation/testing. b) Prediction of excluded datapoints not included in the model training/val/test phase to confirm AGRA is capable of accurate prediction across multiple material systems and adsorbates. A visualization of the newly designed ORR HEA system is provided and highlighted in the parity plot.}
\label{fig:CO2RR_ORR}
\end{figure}

As discussed earlier, one of the benefits of this framework is the ability to combine multiple datasets in order to extrapolate performance metrics on systems not explicitly studied through DFT or experiment. As an example of this application, 5 DFT adsorption energy calculations were performed  through VASP to study the CO2RR HEA systems performance for the ORR. The adsorption energy of O and OH on these HEAs were calculated using the same VASP parameters as the CO2RR dataset and the energy of O and OH in vacuum was obtained as the difference between water molecules energy in vacuum versus OH and H2, similar to the methodology used for the ORR dataset calculations\citep{batchelor2019}. The material systems were composed of the CO2RR datasets HEA system with the ORR datasets adsorbates ( \cref{fig:CO2RR_ORR}). Each HEA system studied was generated using the same neural generator method that Chen utilized \cite{Chen2022}. The 3 GNNs were trained on a curated dataset composed of both ORR and CO2RR datapoints. For each adsorbate dataset (CO,CHO,COOH,O,OH), approximately 200 randomly selected datapoints were combined to generate a 1000 datapoint dataset. The models were then tasked with predicting the 5 never before seen datapoint adsorption energies for comparison against DFT calculations. The ALIGNN model possessed the best average MAE and RMSE scores of 0.068 eV and 0.104 eV (\cref{fig:CO2RR_ORR} a). Although the model was trained on a limited amount of each adsorbate dataset spanning 2 HEA systems, the average RMSE score when tasked with predicting the datapoints excluded from the 1000 point dataset were within 0.02 eV MAE of AGRA models trained on each individual dataset (\cref{fig:CO2RR_ORR} b). This highlights AGRAs potential for dataset concatenation to generate accurate models spanning multiple material systems and chemical reactions. Furthermore, when tasked with predicting the 5 DFT calculated datapoints composed of a new ORR system, AGRA was able to predict the material systems performance trends despite having no prior datapoints on the HEA systems ORR performance. Although AGRA consistently predicted the adsorption energy to be more negative than the DFT calculations by ~0.6 eV, the conclusion that the newly studied ORR catalyst would not be ideal due to strong adsorption of O and OH would still be arrived upon. The 5 DFT calculated datapoints can be found in \cref{fig:CO2RR_ORR}b as the brown points labeled 'New System'. 3 OH adsorbate and 3 O adsorbate datapoints were initially calculated for a total of 6 DFT calculations however 1 of the O adsorbate DFT calculations saw considerable surface migration and was removed as a result. The final 5 datapoints primarily stabilized on hollow sites with only 1 OH adsorbate stabilizing on a bridge site. Notably, the largest prediction errors came from the O adsorbate datapoints.The drop in accuracy can be attributed to the large degree of extrapolation required to perform this type of novel prediction.Aside from AGRA, the furthest extent of extrapolation ML aided catalyst design went to was compositional exploration of a material system and adsorbate combination included in the training data. AGRA takes this one step further by extrapolating to material system and adsorbate combinations never before seen. The benefits of studying pre-established HEA systems for different reactions were proven by Chen with the CO2RR dataset. His work took an already existing catalyst for ammonia decomposition and applied it to the CO2RR to obtain exceptional overpotentials \cite{xie2019}. Since this study can be replicated on any combination of published databases, AGRA has the potential to unify the many separate catalyst databases available to the public to discover unknown applications of pre-established catalysts.

\subsubsection{Pipeline Performance}

As seen in Table II, the ALIGNN model performed the best likely due to its consideration of bond angles with a crystal system as well as the adsorption angle of asymmetric adsorbates such as COOH and OH. Although all 3 models were very close to DFT level accuracy (~0.1 eV) \cite{Zhong2020}, The ALIGNN model was clearly the strongest performing GNN with AGRA.


\begin{table}[!ht]
\caption{\label{tab:performance} Model Performance Summary - Average Mean Absolute Error of 5 model re-trains (in eV)}
    \centering
    \begin{tabular}{ccccc}
    \hline
        \textbf{} &\hspace{0.5cm} \textbf{Graphing Type} &\hspace{0.5cm} \textbf{CGCNN}\hspace{0.5cm} & \textbf{NNConv} & \hspace{0.5cm}\textbf{ALIGNN} \hspace{0.5cm}\\ \hline
        \textbf{CO} & AGRA & 0.140 & 0.084 & 0.095 \\ 
        \textbf{} & OCP & 0.151 & 0.166 & 0.107 \\ 
        \textbf{CHO} & AGRA & 0.099 & 0.102 & 0.095 \\ 
        \textbf{} & OCP & 0.200 & 0.159 & 0.096 \\ 
        \textbf{COOH} & AGRA & 0.065 & 0.100 & 0.065 \\ 
        \textbf{} & OCP & 0.081 & 0.185 & 0.061 \\ 
        \textbf{CO2RR} & AGRA & 0.081 & 0.114 & 0.067 \\ 
        \textbf{} & OCP & 0.177 & 0.165 & 0.091 \\ 
        \textbf{O} & AGRA & 0.123 & 0.136 & 0.047 \\ 
        \textbf{} & OCP & 0.124 & 0.086 & 0.058 \\ 
        \textbf{OH} & AGRA & 0.074 & 0.072 & 0.034 \\ 
        \textbf{} & OCP & 0.062 & 0.052 & 0.038 \\ 
        \textbf{ORR} & AGRA & 0.086 & 0.097 & 0.042 \\ 
        \textbf{} & OCP & 0.083 & 0.094 & 0.048 \\ 
        \textbf{All} & AGRA & 0.300 & 0.260 & 0.068 \\ 
        \textbf{} & OCP & 0.329 & 0.270 & 0.104 \\ \hline
    \end{tabular}
\end{table}

Most notably, the accuracy loss related to combining multiple adsorbate datasets was the least pronounced with the ALIGNN model as well. This shows great potential for extrapolative uses in predicting the performance of material systems for energy reactions without explicitly performing experimental or computational studies. Although composition exploration studies have been performed before, their range of extrapolations has not been extended as far as the study in section 3 which investigated a new adsorbate on a known HEA \cite{Katiyar2021}. \\

When comparing the AGRA to OCP graph representation approaches, it is evident AGRA has similar or superior model performance at a reduced computational cost (Table II). For the individual adsorbate datasets, AGRA performed similar to or superior to the OCP representation but for the datasets which possessed multiple adsorbates, AGRAs representation performed considerably better. This is likely due to the more consistent number of nodes and non fully connected edge representation AGRA provides. 

\section{Conclusion}

In summary, this work has presented an algorithm to analyze and extract the chemical environment of an adsorption site on different metallic substrates. This automated graph representation adds an additional layer of descriptiveness to neural networks which will ultimately allow them to develop greater understandings of the underlying physics of catalysis. This closed system improves model flexibility when combining databases at a reduced computational cost and can accelerate the optimization and discovery process of new catalysts for crucial energy reactions. By removing almost all of the manual curation associated with developing DFT powered datasets for materials design, this package increases the robustness, transferability and explorative capabilities of researchers to allow them to focus on theoretical mechanisms instead of software/technical hurdles to curating datasets. To prove this claim, AGRA was applied to 2 different catalytic reactions to show the exceptional performance of the GNNs on metallic substrates. The ORR dataset obtained 0.048 (*OH) and 0.059 (*O) RMSD on a 871 and 998 datapoint dataset with minimal preparation. To show the versatile learning capabilities of the AGRA, a CO2RR dataset was tested with 691 datapoints split between 3 adsorbates (CHO,CO, COOH) and obtained RMSD values of 0.123,0.125, and 0.093 eV. In each dataset where the OCP representation was benchmarked against AGRA, identical or similar accuracy was obtained for singular adsorbate datasets. For multiple adsorbate datasets however, AGRA performed better while also reducing the computational cost associated with training the GNNs.The dataset combination functionality of AGRA was further highlighted by combining a ORR and CO2RR database to ultimately evaluate a never before studied ORR system to show AGRAs potential to harness the largest publicly available material informatics databases for material design exploration.  
\section*{Supplementary Material}
The Supplementary Materials accompanying the work include the technical limitations of the framework with strategies on how to fix them, visualizations of AGRA's workflow, and hyperparameter's of the models discussed in the work.

\begin{acknowledgments}
The authors acknowledge  financial support from collaborative R\&D programs and initiatives
at National Research Council  through the Artificial Intelligence for Design (AI4D) Challenge program, and the University of Waterloo. They also acknowledge the Nature Science and Engineer Research Council of Canada (NSERC), as well as Digital Research
Alliance of Canada for providing computing resources at the SciNet, CalculQuebec, and Westgrid consortia.
\end{acknowledgments}

\section*{Author Contributions} Conceptualization, Conrard Tetsassi; Implementation and Testing, Conrard Tetsassi and Zachary Gariepy; Manuscript writing, Zachary Gariepy and Conrard Tetsassi; CO2RR dataset generation, Zhiwen Chen, and Zachary Gariepy. All authors discussed the results and contributed to the final manuscript

\section*{Data Availability}
The main datasets that support the findings of this study are openly available in Chen and Batchelor's works at
https://doi.org/10.1021/acscatal.2c03675\citep{Chen2022} and https://doi.org/10.1016/j.joule.2018.12.015\citep{batchelor2019}.  The AGRA code  could be found here \url{https://github.com/Feugmo-Group/AGRA}, as  well as the code used  for the  ML training and figure  plotting.

\section*{Conflicts of Interest} The authors declare no conflict of interest.



\nocite{*}
\bibliography{aipsamp}

\end{document}